\documentclass[10pt, oneside]{article}   	
\usepackage{geometry}                		
\usepackage{graphicx}				
\usepackage{amssymb}
\usepackage{bm}
\title{Context Mixing via Ground State Search\footnote{This work is based on results from a project commissioned by the new Energy and Industrial Technology Development Organization (NEDO), Japan}}
\author{Kentaro Imafuku\\
National Institute of Advanced Industrial Science and Technology (AIST)\\
Aomi 2-3-26, Koto-ku, Tokyo 1350064, Japan}
\date{}							

\begin{document}
\maketitle
\begin{abstract}
To address context mixing problem via ground state search, we introduce an effective Hamiltonian whose ground state presents the best mixing of a prior given probability distributions to approximately describe unknown target probability distribution.
\end{abstract}
\section{Context Mixing}
Context mixing \cite{Hirschberg1992,Mahoney2005,Kulekci2011} is a way to estimate an unknown probability distribution by appropriately mixing a prior given probability distributions. Let 
\begin{equation}
\{p_i(x)\}_{i \in {\mathcal C}}
\end{equation}
 be the given probability distributions where $i\in{\mathcal C}$ indicates a condition that gives context defining $p_i(x)$. For simplicity, we restrict ourselves to the case where $x\in{\mathcal X}$ is discrete. Most intuitive way would be introducing a linear combination as
\begin{equation}\label{eq:linear model}
P_{\bm{\omega}}^{(L)}(x):=\sum_{i\in{\mathcal C}}\omega_i p_i(x)
\end{equation}
with $\omega_i >0$, $\sum_{i \in {\mathcal C}}\omega_i=1$ and $\bm{\omega}:=\{\omega_i\}_{i\in{\mathcal C}}$. Under the parameterization, context mixing takes form of finding parameter $\bm{\omega}^*:=\{\omega_i^*\}_{i\in{\mathcal C}}$ such that $P^{(L)}_{\bm{\omega}\rightarrow \bm{\omega}^*}(x)$ becomes a ``good" approximation of a target probability distribution $\mu(x)$. Besides the parameterization in eq.(\ref{eq:linear model}), another parameterization as
\begin{equation}\label{eq:maximum entropy parameterization}
P^{(ME)}_{\bm{\omega}}(x)=\frac{1}{Z(\bm{\omega})}\exp\left(\sum_{i\in{\mathcal C}}\omega_i \eta_i(x)\right)
\end{equation}
with
\begin{equation}
\eta_i(x):=-\ln p_i(x),\quad\mbox{and}\quad
Z(\bm{\omega})=\sum_{x\in {\mathcal X}} \exp\left(\sum_{i\in{\mathcal C}}\omega_i \eta_i(x)\right)
\end{equation}
is also often considered\cite{Mahoney2005}. The form of parameterization in eq.(\ref{eq:maximum entropy parameterization}) which is derived based on the maximum entropy principle\cite{PhysRev.106.620,jaynes_justice_1986} guarantees that $P^{(ME)}_{\bm{\omega}}(x)$ is the probability distribution with the maximum entropy among others which have the same expectation values with
\begin{equation}
\tau_i:=\sum_{x\in {\mathcal X}}\eta_i(x) P^{(ME)}_{\bm{\omega}}(x)=
-\sum_{x\in {\mathcal X}}P^{(ME)}_{\bm{\omega}}(x) \ln p_i(x).
\end{equation}
Notice that $\tau_i$ corresponds to the mean code length in the case where $x$ following $P^{(ME)}_{\bm{\omega}}(x)$ is coded into an entropic code based on $p_i(x)$\cite{6773067}. ($x$ is encoded into a code word whose length is $\ln p_i^{-1}(x)$.) In addition, extra length in comparison with the case where $x$ is naturally coded into the entropic code based on $P^{(ME)}_{\bm{\omega}}(x)$ is given as
\begin{eqnarray}
\Delta \tau_i :&=&\tau_i-\left(-\sum_{x\in {\mathcal X}}P^{(ME)}_{\bm{\omega}}(x) \ln P^{(ME)}_{\bm{\omega}}(x)\right)
\\
&=&\sum_{x\in {\mathcal X}}P^{(ME)}_{\bm{\omega}}(x)\ln \frac{P^{(ME)}_{\bm{\omega}}(x)}{p_i(x)}
\end{eqnarray}
that is Kullback-Leibler distance between $P^{(ME)}_{\bm{\omega}}(x)$ and $p_i(x)$. Thus, choosing the maximum entropy state with respect to $\tau_j$ implies minimizing $\Delta \tau_j$, so as any extra conditions except for $\tau_j$ can be excluded in constructing $P^{(ME)}_{\bm{\omega}}(x)$. (The idea may be similar to a situation where an average drawing is employed except for particularly suggested marks by witnesses in making a facial composite sketch, or principle of the Occam's razor\cite{nla.cat-vn3538300}.)

Kullback-Leibler distance\cite{kullback1951} between $\mu(x)$ and $P^{(ME)}_{\bm{\omega}}(x)$, i.e.,
\begin{equation}
D(\bm{\omega}):=
\sum_{x\in {\mathcal X}}\mu(x)\ln \frac{\mu(x)}{P^{(ME)}_{\bm{\omega}}(x)}
\end{equation}
 can be adopted as a metric in choosing parameter $\bm{\omega}$. In other words, our aim is to find $\bm{\omega}$ minimizing $D(\omega)$, or equivalently minimizing
 \begin{equation}\label{eq:E(w)}
E(\bm{\omega}):=-\sum_{x\in {\mathcal X}}\mu(x) \ln P^{(ME)}_{\bm{\omega}}(x)
 \end{equation}
for the (unknown) target probability distribution $\mu(x)$.

\section{Mapping to Ground State Search}
In the following, we investigate a way to find $\bm{\omega}$ as a solution of the ground state search of a Hamiltonian. We assume that a mixed state
\begin{equation}\label{eq:hat_mu}
\hat{\mu}:=\sum_{x\in{\mathcal X}}\mu(x) |x\rangle\langle x|
\end{equation}
is physically available although it does not mean that we know $\hat{\mu}$. Roughly speaking, our aim is to construct Hamiltonian $\hat{H}$ such that
\begin{equation}\label{eq:rough_idea}
\hat{H}|\bm{\omega}\rangle=E(\bm{\omega})|\bm{\omega}\rangle
\end{equation}
with $E(\bm{\omega})$ given in eq.(\ref{eq:E(w)}). With the formulation, a ground state of $\hat{H}$ obviously gives the solution $\omega$ minimizing eq.(\ref{eq:E(w)}). In the following, we assume that $\bm{\omega}=\{\omega_i\}_{i \in {\mathcal C}}$ is discrete, i.e.,
\begin{equation}
\omega_i \in \Omega_i
\end{equation}
with a finite set $\Omega_i$, although $\omega_i$ was supposed continuous in the previous section. By appropriately choosing $\Omega_i$, we can obtain an approximate solution that is practically enough in most cases.

Rewriting $P^{(ME)}_{\bm{\omega}}(x)$ in eq.(\ref{eq:maximum entropy parameterization}) as
\begin{equation}
P^{(ME)}_{\bm{\omega}}(x)=\exp\left(\lambda(\bm{\omega})+\sum_{i\in{\mathcal C}} \omega_i \eta_i(x)\right)
\end{equation}
with
\begin{equation}\label{eq:constraint}
\lambda(\bm{\omega})=-\ln Z(\bm{\omega}),
\end{equation}
substituting it into eq.(\ref{eq:E(w)}), we obtain
\begin{equation}
E(\bm{\omega})=-\sum_{x\in {\mathcal X}}\mu(x)\left(\lambda(\bm{\omega})+\sum_{i\in{\mathcal C}}\omega_i\eta_i(x)\right)
\end{equation}
Setting the constraint described in eq.(\ref{eq:constraint}) aside, let us introduce a parameter $\lambda \in \Lambda$ and a function
\begin{equation}
E'(\lambda,\bm{\omega}):=-\sum_{x\in {\mathcal X}}\mu(x)\left(\lambda+\sum_{i\in{\mathcal C}}\omega_i\eta_i(x)\right).
 \end{equation}
Similar to eq.(\ref{eq:rough_idea}), we can introduce a Hamiltonian and its eigenstates as
\begin{equation}
\hat{H}'|\lambda,\bm{\omega}\rangle=E'(\lambda,\bm{\omega})|\lambda,\bm{\omega}\rangle.
\end{equation}
Notice, however, that the ground state of $\hat{H}'$ does not present $\bm{\omega}$ minimizing $E(\bm{\omega})$ in eq.(\ref{eq:E(w)}), unlike the rough idea illustrated in eq.(\ref{eq:rough_idea}). In fact, $E'(\lambda,\bm{\omega})$ is a monotonic decreasing function with respect to $\lambda$, and is not lower bounded. To address this point, adding a constraint term to $E'(\lambda,\bm{\omega})$, we introduce
\begin{equation}
E(\lambda,\bm{\omega})=E'(\lambda,\bm{\omega})+\alpha \left(\sum_{x\in {\mathcal X}}\exp\left(\lambda+\sum_{i\in{\mathcal C}}\omega_i\eta_i(x)\right)-1\right)^2
\end{equation}
with $\alpha >0$. By constructing a Hamiltonian such as
\begin{equation}\label{eq:energy_function2}
\hat{H}|\lambda,\bm{\omega}\rangle=E(\lambda,\bm{\omega})|\lambda,\bm{\omega}\rangle,
\end{equation}
we can safely reduce the context mixing problem to the ground state search of the Hamiltonian. To construct it, we consider a grand Hamiltonian

\begin{eqnarray}
\hat{G}:\!\!\!\!&=&\!\!\!\!-\sum_{x\in {\mathcal X}}
\left(\hat{\lambda}+\sum_{i\in {\mathcal C}}\hat{\omega}_i\eta_i(x)\right)\otimes|x\rangle\langle x|\nonumber\\
&&\!\!\!\!
+\alpha\left(|{\mathcal X}|^2\sum_{x'\in {\mathcal X}}\sum_{x''\in {\mathcal X}}
e^{\hat{\lambda}+\sum_{i\in{\mathcal C}}\hat{\omega}_i\eta_i(x')}
e^{\hat{\lambda}+\sum_{i\in{\mathcal C}}\hat{\omega}_i\eta_i(x'')}
\otimes
|x'\rangle\langle x'|\otimes|x''\rangle\langle x''|\right.\nonumber\\
&&\left.
~~~~~-2|{\mathcal X}|\sum_{x'\in {\mathcal X}}
e^{\hat{\lambda}+\sum_{i\in{\mathcal C}}\hat{\omega}_i\eta_i(x')}
\otimes
|x'\rangle\langle x'|
+\hat{I}\right)
\label{eq:grand Hamiltonian}
\end{eqnarray}
where
\begin{equation}
\hat{\lambda}:=\sum_{\lambda\in \Lambda}\lambda|\lambda\rangle\langle\lambda|,\quad
\hat{\omega}_i:=\sum_{\omega_i\in \Omega_i}\omega_i|\omega_i\rangle\langle\omega_i|
\end{equation}
with state vectors
\begin{equation}
|\lambda\rangle\in {\mathcal H}_\Lambda,\quad
|\omega_i\rangle\in {\mathcal H}_{\Omega_i},\quad
|x\rangle\in {\mathcal H}_{{\mathcal X}_0},\quad
|x'\rangle\in {\mathcal H}_{{\mathcal X}_1},\quad\mbox{and,}\quad
|x''\rangle\in {\mathcal H}_{{\mathcal X}_2}.
\end{equation}
Notice that the full Hilbert space where $\hat{G}$ lives is 
\begin{equation}\label{eq:grand Hilbert space}
{\mathcal H}:={\mathcal H}_{\Lambda}\otimes{\mathcal H}_\Omega\otimes{\mathcal H}_{{\mathcal X}_0}\otimes{\mathcal H}_{{\mathcal X}_1}\otimes{\mathcal H}_{{\mathcal X}_2}
\end{equation}
where
\begin{equation}
{\mathcal H}_\Omega:=\bigotimes_{i\in {\mathcal C}}{\mathcal H}_{\Omega_i}.
\end{equation}
Now, let us suppose that a state on ${\mathcal H}_{{\mathcal X}_0}\otimes{\mathcal H}_{{\mathcal X}_1}\otimes{\mathcal H}_{{\mathcal X}_2}$ is somehow fixed into a state
\begin{equation}\label{eq:state nu}
\hat{\nu}:=\hat{\mu}\otimes\frac{\hat{I}}{|{\mathcal X}|}\otimes\frac{\hat{I}}{|{\mathcal X}|}
\end{equation}
where $\hat{\mu}$ is given in eq.(\ref{eq:hat_mu}), and $I/|{\mathcal X}|$ is the complete mixed state on each Hilbert space.
In this case, $\hat{G}$ in eq.(\ref{eq:grand Hamiltonian}) behaves on partial Hilbert space ${\mathcal H}_{\Lambda}\otimes{\mathcal H}_{\Omega}$ as an effective Hamiltonian described as
\begin{eqnarray}\label{eq:H_eff}
\hat{H}_{eff}:\!\!\!\!&=&\!\!\!\!{\rm tr}_{{\mathcal H}_{{\mathcal X}_0}\otimes{\mathcal H}_{{\mathcal X}_1}\otimes{\mathcal H}_{{\mathcal X}_2}}\left[\hat{\nu}~\hat{G}\right]\nonumber\\
&=&\!\!\!\!E'(\hat{\lambda},\hat{\bm{\omega}})+\alpha \left(\sum_{x\in {\mathcal X}}\exp\left(\hat{\lambda}+\sum_{i\in{\mathcal C}}\hat{\omega_i}\eta_i(x)\right)-\hat{I}\right)^2
\end{eqnarray}
that is nothing but $\hat{H}$ in eq.(\ref{eq:energy_function2}). 

Let us consider an application of $\hat{H}_{eff}$ to the quantum annealing computation\cite{PhysRevE.58.5355,2000quant.ph..1106F,Farhi2000ANS,RevModPhys.80.1061,2014McGeoch} with a driving Hamiltonian $\hat{V}$ on ${\mathcal H}_\Lambda\otimes{\mathcal H}_\Omega$, introducing a time dependent Hamiltonian such as
\begin{equation}
\hat{H}_{eff}(t)=\frac{t}{T}\hat{H}_{eff}+\left(1-\frac{t}{T}\right)\hat{V}
\end{equation}
with initial condition
\begin{equation}
|\varphi_0\rangle \in {\mathcal H}_\Lambda\otimes{\mathcal H}_\Omega
\end{equation}
that is chosen to be the ground state of $\hat{V}$. As a dynamics on the grand Hilbert space in eq.(\ref{eq:grand Hilbert space}), the above is realized by introducing a time dependent Hamiltonian such as
\begin{equation}
\hat{G}(t)=\frac{t}{T}\hat{G}+\left(1-\frac{t}{T}\right)\hat{V}\otimes \hat{I}
\end{equation}
where $\hat{I}$ is the identity operator on ${\mathcal H}_{{\mathcal X}_0}\otimes{\mathcal H}_{{\mathcal X}_1}\otimes{\mathcal H}_{{\mathcal X}_2}$, with the initial state
\begin{equation}\label{eq:product form}
\hat{\rho}_0:=|\varphi_0\rangle\langle\varphi_0|\otimes\hat{\nu}.
\end{equation}
Notice that the state evolution by $\hat{G}(t)$ {\it does not} generally preserve the product form in eq.(\ref{eq:product form}). Moreover,
since
${\rm tr}_{{\mathcal H}_{{\mathcal X}_0},{\mathcal H}_{{\mathcal X}_1},{\mathcal H}_{{\mathcal X}_2}}\left(\hat{G}^2\right)$ is not $\hat{H}_{eff}^2$ on Hilbert space ${\mathcal H}_\Lambda\otimes{\mathcal H}_{\Omega}$, the dynamics of the reduced state into the Hilbert space governed by $\hat{G}(t)$ deviates from the dynamics
\begin{equation}\label{eq:original dynamics}
\frac{d}{dt}|\varphi_t\rangle=-i \hat{H}_{eff}(t)|\varphi_t\rangle
\end{equation}
in the order of $O(\delta t^2)$ for small time interval $\delta t$. To resolve this point, we need to supply new $\hat{\nu}$ one after another after every short time evolutions by time interval $\delta t$. By this procedure, we can simulate the dynamics in eq.(\ref{eq:original dynamics}) for a finite time interval with error up to the order of $O(\delta t)$.

\end{document}